%% file: KOTLARSKI_Wojciech_DISCRETE2016.tex
\documentclass[a4paper]{jpconf}

\usepackage{graphicx}
\usepackage[utf8]{inputenc}
\usepackage{amsmath}
\usepackage{amssymb}
\DeclareMathOperator{\arctanh}{arctanh}
\usepackage[sort&compress,square,numbers]{natbib}
\usepackage{subfig}
\begin{document}
\title{Scalar color octets and triplets in the SUSY with R-symmetry}

\author{Wojciech Kotlarski}

\address{
    Faculty of Physics, University of Warsaw, Pasteura 5, 02-093 Warsaw, Poland \\
    Institute for Nuclear and Particle Physics, TU Dresden, Zellescher Weg 19, 01069 Dresden, Germany
}

\ead{wojciech.kotlarski@fuw.edu.pl}

\input{abstract.tex}
\input{introduction.tex}
\input{model.tex}
\input{sgluon.tex}
\input{squark.tex}
\input{conclusions.tex}

\input{acknowledgements.tex}

\appendix
\section{}
\subsection{BMP1}

\begin{verbatim}

Results for subprocess uu > suLsuR(+X)
---------------------------------------------------------------
       tree:  5.67655e+00 +/- 5.7e-06 fb ( p-value =  0.0e+00 )
    virtual:  3.53574e+00 +/- 3.5e-05 fb ( p-value =  9.9e-17 )
real (soft): -5.71007e+01 +/- 4.4e-05 fb ( p-value =  0.0e+00 )
real (hard):  5.78352e+01 +/- 5.8e-04 fb ( p-value =  0.0e+00 )
---------------------------------------------------------------
        sum:  9.94681e+00 +/- 5.8e-04 fb

Results for subprocess gu > suLsuR(+X)
---------------------------------------------------------------
       tree:  0.00000e+00 +/- 0.0e+00 fb ( p-value =  0.0e+00 )
    virtual:  0.00000e+00 +/- 0.0e+00 fb ( p-value =  0.0e+00 )
real (soft): -5.03323e-02 +/- 5.0e-09 fb ( p-value =  0.0e+00 )
real (hard):  4.66389e-02 +/- 4.7e-07 fb ( p-value =  0.0e+00 )
---------------------------------------------------------------
        sum: -3.69334e-03 +/- 4.7e-07 fb

Results for subprocess sum
---------------------------------------------------------------
       tree:  5.67655e+00 +/- 5.7e-06 fb ( p-value =  0.0e+00 )
    virtual:  3.53574e+00 +/- 3.5e-05 fb ( p-value =  9.9e-17 )
real (soft): -5.71511e+01 +/- 4.4e-05 fb ( p-value =  0.0e+00 )
real (hard):  5.78819e+01 +/- 5.8e-04 fb ( p-value =  0.0e+00 )
---------------------------------------------------------------
        sum:  9.94312e+00 +/- 5.8e-04 fb  
\end{verbatim}

\subsection{BMP2}
\begin{verbatim}

Results for subprocess uu > suLsuR(+X)
---------------------------------------------------------------
       tree:  2.01180e+00 +/- 2.0e-06 fb ( p-value =  0.0e+00 )
    virtual:  6.97868e-01 +/- 7.0e-06 fb ( p-value =  0.0e+00 )
real (soft): -1.99882e+01 +/- 1.6e-05 fb ( p-value =  0.0e+00 )
real (hard):  2.02943e+01 +/- 2.0e-04 fb ( p-value =  0.0e+00 )
---------------------------------------------------------------
        sum:  3.01582e+00 +/- 2.0e-04 fb
        
Results for subprocess gu > suLsuR(+X)
---------------------------------------------------------------
       tree:  0.00000e+00 +/- 0.0e+00 fb ( p-value =  0.0e+00 )
    virtual:  0.00000e+00 +/- 0.0e+00 fb ( p-value =  0.0e+00 )
real (soft): -1.69119e-02 +/- 1.7e-09 fb ( p-value =  0.0e+00 )
real (hard):  1.46572e-02 +/- 1.5e-07 fb ( p-value =  0.0e+00 )
---------------------------------------------------------------
        sum: -2.25477e-03 +/- 1.5e-07 fb

Results for subprocess sum
---------------------------------------------------------------
       tree:  2.01180e+00 +/- 2.0e-06 fb ( p-value =  0.0e+00 )
    virtual:  6.97868e-01 +/- 7.0e-06 fb ( p-value =  0.0e+00 )
real (soft): -2.00051e+01 +/- 1.6e-05 fb ( p-value =  0.0e+00 )
real (hard):  2.03090e+01 +/- 2.0e-04 fb ( p-value =  0.0e+00 )
---------------------------------------------------------------
        sum:  3.01357e+00 +/- 2.0e-04 fb
\end{verbatim}

\bibliographystyle{iopart-num}
\bibliography{bibliography.bib}

\end{document}

%% file: abstract.tex
\begin{abstract}
In this note we report on the recent progress in the study of the strongly interacting sector of the Minimal R-symmetric Supersymmetric Standard Model (MRSSM).
First, we discuss the limits originating from the search of the sgluon pair production in the same-sign lepton final state at the LHC.
Next, we present the first available in literature calculation of the NLO SQCD corrections to the same-sign squark pair production in the MRSSM.
We discuss the relative size of $k$-factors compared to the MSSM and some technical details that led to this result. 
\end{abstract}

%% file: introduction.tex
\section{Introduction}

The Minimal Supersymmetric Standard Model (MSSM), with its simple structure and rich phenomenology, has been for many years the primary realization of supersymmetry considered in the literature.
But as the Large Hadron Collider continues its second run with no MSSM in sight, the question is being raised whether low-scale supersymmetry exists at all.
While it is possible that this is indeed not the case, it is also possible that supersymmetry is just not realized in the minimal, MSSM way.
This is the reason behind increased attention in the non-minimal SUSY models observed in recent years.

One of the models which proved to be very promising in this context is the Minimal R-symmetric Supersymmetric Standard Model (MRSSM) \cite{Kribs:2007ac}.
The model ameliorates the flavor problem of the MSSM \cite{Kribs:2007ac,Dudas:2013gga} while reducing fine-tuning \cite{Bertuzzo:2014bwa} and exhibiting distinct, interesting phenomenology. 
Though very promising, it is also much less explored - with the MSSM having a 30-year head start.
Until now, the main effort has been devoted to the study of its electroweak sector, especially in the context of the discovery of the Higgs boson \cite{Diessner:2014ksa,Bertuzzo:2014bwa,Diessner:2015yna} and studies of dark matter \cite{Diessner:2015iln}.
The phenomenology of the strongly interacting sector, which is the most important from the point of view of direct detection at the LHC, was much less investigated.
The focus there has so far been mainly on the sgluon production and LO analyses \cite{Choi:2008ub,Plehn:2008ae,Calvet:2012rk,Kotlarski:2013lja,Kotlarski:2016zhv,Choi:2008pi,Kribs:2013oda}.\footnote{Strictly speaking, some of these works do not deal with the MRSSM but with a model in which $\mathcal{N}=1$ SUSY is extended in the strongly interacting sector by $\mathcal{N}=2$ gauge companions, which does have a very similar SQCD sector.}
As the super-QCD (SQCD) $k$-factors tend to be large, a question arises of how the latter studies change in the presence of quantum corrections.
To fill this gap, in this work, after the recap of the recent results of sgluon analysis, we report on the progress in calculation of the next-to-leading order (NLO) SQCD corrections to the same-sign squark pair production.

The paper is structured as follows.
In section 2, we briefly recap the main features of the MRSSM.
In section 3, we discuss the phenomenology of the color octet scalars (sgluons) production at the 13 TeV LHC.
Finally, in section 4, we report on the calculation of the NLO SQCD corrections to the same-sign squark pair production.
To that end, we discuss the calculation of virtual corrections together with their renormalization and derivation of the SUSY-restoring counterterms.
We also describe the treatment of  soft and collinear divergences and resonances appearing in real corrections.
We end with a comparison of cross sections between the MSSM and the MRSSM for 2 benchmark points with different squark-gluino mass hierarchies.

%% file: model.tex
\section{The Model}
The guiding principle behind the construction of the MRSSM is the presence of the unbroken R-symmetry at the low scale.
However, with the R-charge assignment mimicking R-parity, the construction of phenomenologically viable model does require the extension of the MSSM field content \cite{Kribs:2007ac}.
In the strongly interacting sector this amounts to adding a color-octet (electroweak singlet) chiral superfield $\hat O$, whose fermionic component allows one to construct a Dirac mass term for the Dirac gluino $\tilde g_D$,
\begin{equation}
\tilde g_D \equiv \left (
  \begin{matrix}
    \tilde g \\
    \overline{\tilde O}
  \end{matrix}
\right ),
\end{equation}
\begin{table} 
\begin{center}
\caption{
    Strongly interacting superfield content of the MRSSM together with R-charges of the component fields. 
    The superfield in the last line is absent in the MSSM. 
    It comprises of the right-handed component of the Dirac gluino and two real scalar gluons.
    \label{tab:fieldcontent} \\
}
\begin{tabular}{c c||c|c||c|c}
\multicolumn{2}{c||}{superfield} & \multicolumn{2}{c||}{boson} & \multicolumn{2}{c}{fermion} \\
\hline 
 left-handed (s)quark &$\hat{Q}_L$ &  $\tilde{q}_L$ &  $1$ & $q_L$ & $0$\\
 right-handed (s)quark &$\hat{Q}_R$ &  $\tilde{q}_R^\dagger$ &  $1$ &  $\bar q_R$ &  $0$\\
 gluon vector superfield &$\hat{g}$ &  $g$ & 0 &  $\tilde{g}$ & $+1$\\ 
 adjoint chiral superfield &$\hat{O}$ & $O$ &  $0$ & $\tilde{O}$ & $-1$
\end{tabular}
\end{center}
\end{table}
where, as shown in table~\ref{tab:fieldcontent}, $\tilde g$ and $\tilde O$ are the gluino and the octino Weyl fermions, respectively.

The addition of the superfield $\hat O$ has important phenomenological consequences.
For one, its scalar, CP-odd component $O_A$ can be arbitrarily light as its tree-level mass is controlled purely by the soft-breaking parameter.
This makes it ideal to search for at the LHC, also in the framework of simplified models.
Also, heavier squark pair production is dominated by same-sign squarks, a process mediated by the exchange of a gluino in the t-channel.
In the MSSM, due to gluino's Majorana nature, it is possible to produce $\tilde q_L \tilde q_L$, $\tilde q_R \tilde q_R$ and $\tilde q_L \tilde q_R$ final states while in the MRSSM only the $\tilde q_L \tilde q_R$ combination is allowed.
This lowers the inclusive squark production cross section, weakening the LHC exclusion limits.
This comes at the expense of the gluino mass limits, which are more stringent for Dirac gluinos.

\begin{figure}
	\centering
    \includegraphics[width=0.99\textwidth]{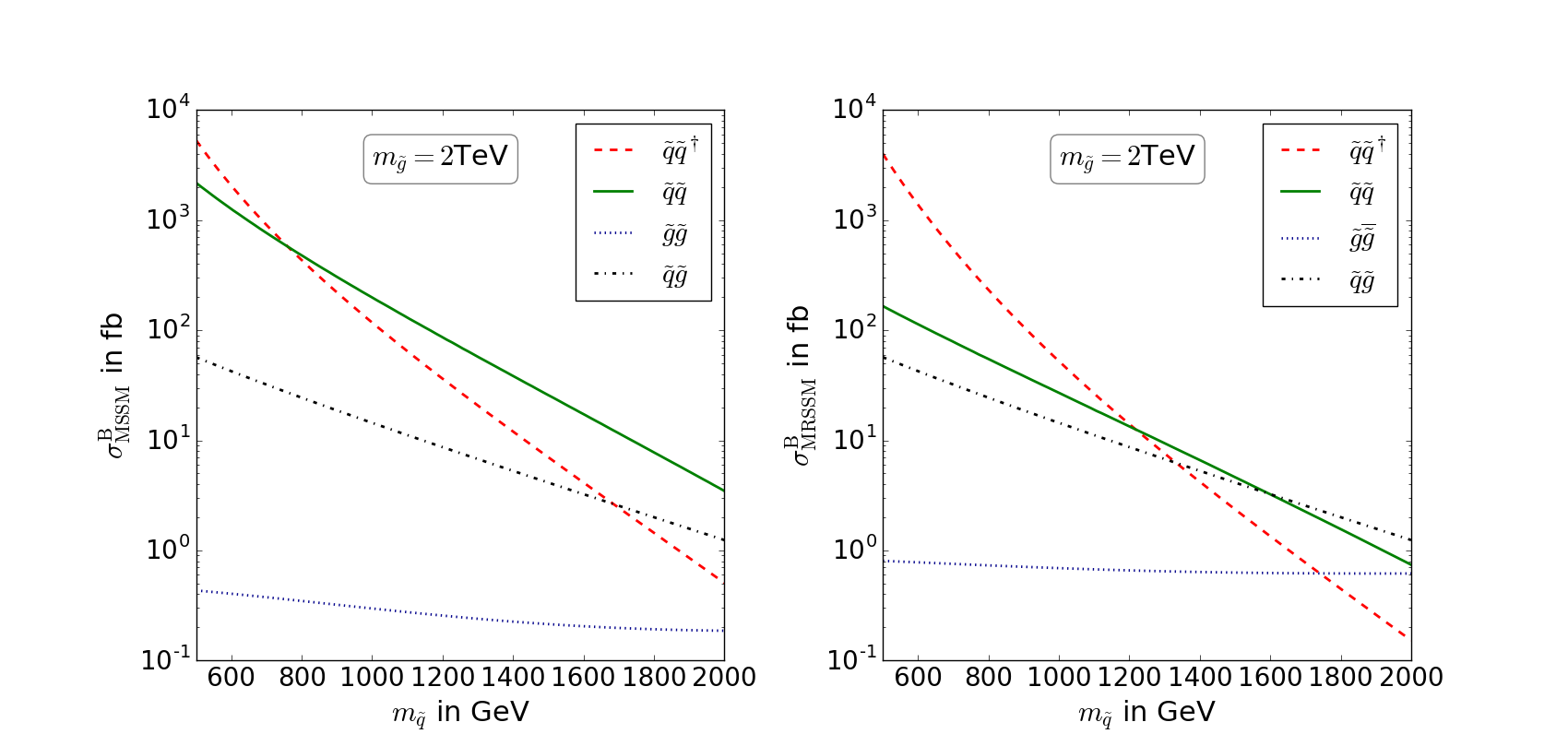}
	\caption{
        Comparison of the selected cross sections between the MSSM and the MRSSM in function of the squark mass. 
        Different quark flavors (excluding stops) and ``chiralities" are added together.
    \label{fig:xsec}}
\end{figure}

To illustrate this discussion, in figure \ref{fig:xsec} we compare the MRSSM and MSSM cross sections for the pair production of various strongly interacting SUSY particles present in these models.
In the plot we use the \texttt{MMHT2014lo68cl}~\cite{Harland-Lang:2014zoa} PDFs interfaced through \texttt{LHAPDF6}~\cite{Buckley:2014ana} with factorization and renormalization scales set equal to arithmetic mean of the masses of final-state particles.        
As can be seen, assuming the gluino mass of 2 TeV, for masses of squarks of around 1 TeV the MSSM cross section is dominated by the (inclusive) same-sign squark pair production.
As stated in the previous paragraph, the analogous cross section in the MRSSM is much smaller while the opposite-sign squark production cross section is quite similar to the MSSM.
The sgluon pair production cross section, which is an exclusive feature of the MRSSM, is also of the same order.

Since we are primarily interested in the differences between the MSSM and the MRSSM, we will focus our discussion in the next two sections on production of sgluon pairs and same-sign squarks.

%% file: sgluon.tex
\section{Sgluon pair production at the 13 TeV LHC}
The partonic cross sections for the pair production of scalar or pseudoscalar sgluons are given by
\begin{align}
  \label{eq:sgluon_xsec1}
  \hat{\sigma} (q \bar{q} \to O O)  = & \frac{2\pi\alpha_s^2}{9\hat{s}} \beta^3_O ,\\
  \label{eq:sgluon_xsec2}
  \hat{\sigma} (g g \to O O) = & \frac{3 \pi \alpha_s^2}{32 \hat{s}} \left ( 27 \beta_O - 17 \beta^3_O +6 (-3 + 2\beta^2_O + \beta^4_O )\arctanh \beta_O \right  ),  
\end{align}
where $\hat s$ is the standard partonic Mandelstam variable and $\beta_O \equiv \sqrt{1 - 4 m_O^2/\hat s}$ is the sgluon's velocity in the center-of-mass system of colliding partons.
\begin{figure}
  \centering
  \includegraphics{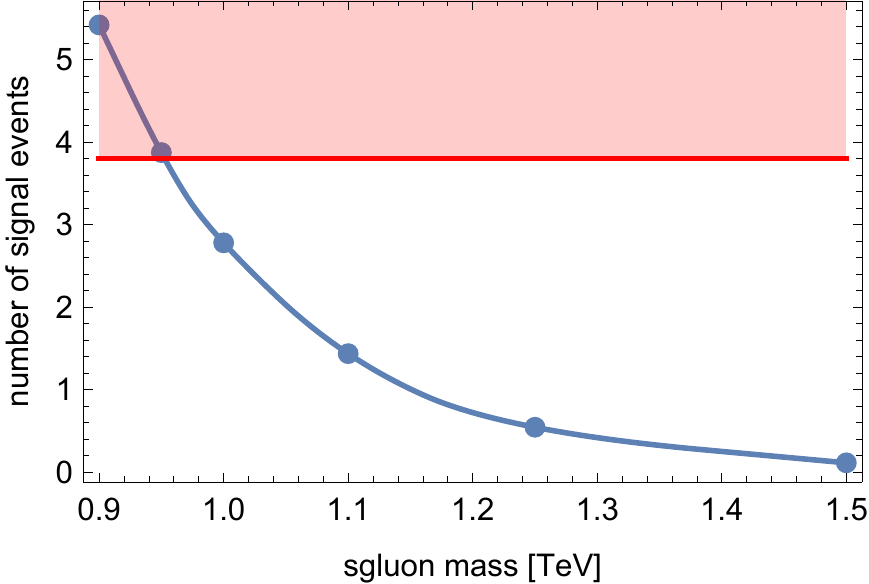}
  \caption{Predicted number of observed signal events as a function of the sgluon mass (blue points).
    Solid line shows interpolation between these points.
    Red region is excluded by ATLAS for SR3b of \cite{Aad:2016tuk} at 95\% CL.
    Interpreted in the context of sgluon production, it corresponds to a lower limit on the sgluon mass $m_O \gtrsim 0.95$ TeV.
    \label{fig:sgluon_mass_exclusion}}
\end{figure}
In the MRSSM, the masses of the scalar ($S$) and the pseudoscalar ($A$) components of the field $O$ from table \ref{tab:fieldcontent}, where $O \equiv {\textstyle\frac{1}{2}} (O_S + \imath O_A)$,  are related  through the gluino mass $M_O^D$ (in the normalization used in Ref.~\cite{Diessner:2014ksa}) as 
\begin{align}
\label{eq:mass_relation}
m_{O_S}^2 = m_{O_A}^2 + 4 (M_O^D)^2.
\end{align} 
Since LHC constraints require gluinos to be $\gtrsim 1.5$ TeV, light scalar sgluon would require a very light pseudoscalar, which is highly constrained.
For phenomenological analysis, we therefore focus on the scenario where $m_{O_A} \simeq 1$ TeV.

Once produced, pseudoscalar sgluons will decay almost exclusively to top quark pairs (see Appendix E of \cite{Kotlarski:2016nop}).
This produces a 4-top quark final state, a signature that is rare in the SM.
Using the search of the same-sign lepton production performed by ATLAS~\cite{Aad:2016tuk}, we have constrained the sgluons' contribution to this signature \cite{Kotlarski:2016zhv}.
This, as shown in figure \ref{fig:sgluon_mass_exclusion}, directly translates to the lower limit on the pseudoscalar sgluon mass of 0.95 TeV.

Possibility of the existence of a relatively light color-charged particle that decays directly to standard model particles is one of the distinguishing features of models with Dirac gluinos.
If ever discovered, it will be a clear falsification of the MSSM.

%% file: squark.tex
\section{Same-sign squark pair production at the NLO}
The second important channel is the same-sign squark pair production.
The partonic cross section for the pair production of left and right squarks of flavor $i$ and equal masses in the MSSM and in the MRSSM is given by
\begin{align}
\hat \sigma(q_i q_i \to \tilde{q}_{i,L} \tilde{q}_{i,R}) = \frac{\pi \alpha_s^2}{\hat s} \Bigg[ & 
-\frac{8}{9}\sqrt{1-\frac{4 m_{\tilde{q}_i}^2}{\hat s}} +  \left( -\frac{4}{9} - \frac{8(m_{\tilde{g}}^2 - m_{\tilde{q}_i}^2)}{9 \hat s} \right) \\
 & \left. \times \ln \frac{\hat s \left(1 - \sqrt{1-\frac{4 m_{\tilde{q}_i}^2}{\hat s}}\right) +2 (m_{\tilde{g}}^2 - m_{\tilde{q}_i}^2)}{\hat s \left(1 +  \sqrt{1-\frac{4 m_{\tilde{q}_i}^2}{\hat s}}\right) +2 (m_{\tilde{g}}^2 - m_{\tilde{q}_i}^2)} \right] \nonumber,
\end{align}
where $m_{\tilde g}$ is (Dirac or Majorana) gluino mass and $\hat s$ is as in eqs \ref{eq:sgluon_xsec1} and \ref{eq:sgluon_xsec2}. 
As noted in section 2, in the MSSM one also has additional channels for left-left and right-right squark pair production.
While these tree-level results are enough for qualitative discussions as done in figure \ref{fig:xsec}, a detailed comparison of the MSSM and the MRSSM does require going beyond that.
We therefore present here the calculation of NLO SQCD corrections to the left-right same-sign squark pair production in the MRSSM.

The setup of the calculation is as follows.
We created a custom \texttt{FeynArts}~\cite{Hahn:2000kx} MRSSM model file containing UV and DREG (dimensional regularization) to DRED (dimensional reduction)~\cite{Siegel79,Capper79} transition countertems. 
Both the real and virtual amplitudes were generated using \texttt{FeynArts} and evaluated in \texttt{FormCalc}~\cite{Hahn:1998yk}. 
The amplitudes were then processed in \textit{mathematica} and exported to a standalone \texttt{C++} code.
\subsection{Virtual corrections}
The evaluation of the virtual amplitude follows standard procedures, so here we only outline the most important details.
The amplitude is regularized using the so-called 't Hooft-Veltman scheme (HV in the notation of Ref.~\cite{Signer:2008va}).
In the HV, internal gluons are kept D-dimensional, which breaks supersymmetry due to the mismatch between gluon and gluino number of degrees of freedom.
This is corrected by supplementing the result with a set of counterterms that translate the UV-poles from DREG to DRED, as DRED was proven to not break SUSY at the one-loop level (for reviews of checks see
e.g. \cite{Jack:1997sr,Stockinger:2005gx}).
For the process at hand, only the renormalisation constant of the squark-quark-gluino vertex needs to be changed. 
Denoting the renormalization constant of this vertex by $\delta \hat{g}_s$, we find that it has to satisfy
\begin{align}
\delta \hat{g}_s & = \delta g_s + \frac{g_s^3}{16\pi^2}\left( \frac{2 C_A}{3} - \frac{C_F}{2} \right),
\end{align}
where $\delta g_s$ is the strong coupling constant counterterm in DREG and zero momentum subtraction scheme (more precisely, we use zero momentum subtraction for contributions from particles heavier than the bottom quark and $\overline{\text{MS}}$ for contributions from bottom and light quarks).
The remaining counterterms are fixed by the on-shell renormalization conditions.

After this procedure the amplitude is UV finite and supersymmetric but it still contains the infrared (IR) singularities.

\subsection{Real corrections}
The IR singularities of the virtual amplitude cancel after adding real corrections - soft ones cancel according to the Kinoshita-Lee-Nauenberg theorem \cite{Kinoshita:1962ur,Lee:1964is} while collinear ones are removed through mass factorization \cite{Collins:1985ue,Bodwin:1984hc}.
To extract soft and collinear divergences, we employed the two-cut phase space slicing method as documented in \cite{Harris:2001sx}.
In short, this amounts to splitting the real emission contribution phase space according to emitted gluon energy into soft region defined as
\begin{align}
  \label{eq:soft_region}
  E_g \leq \delta_s \frac{\sqrt{\hat s}}{2},
\end{align}
where $\delta_s \ll 1$, and its complement.
This region of phase space does contain both soft and/or collinear divergences.
Using the eikonal approximation, the divergences can be extracted in D-dimensions as poles in 4-D.

However, the complement of the phase space defined by eq.~\ref{eq:soft_region} still contains collinear divergences.
These are extracted by slicing the phase space according to collinearity condition.
We use the collinear condition from \cite{Dawson:2006dm} as it allows, contrary to the one used in \cite{Harris:2001sx}, to decouple the soft and collinear limits.

The IR finiteness of the result was checked by evaluating squared matrix elements for a single phase space point, showing cancellation of divergences at the level of a double numerical precision.

\subsection{Treatment of on-shell resonances in real emission diagrams}
\begin{figure}
  \centering
  \includegraphics{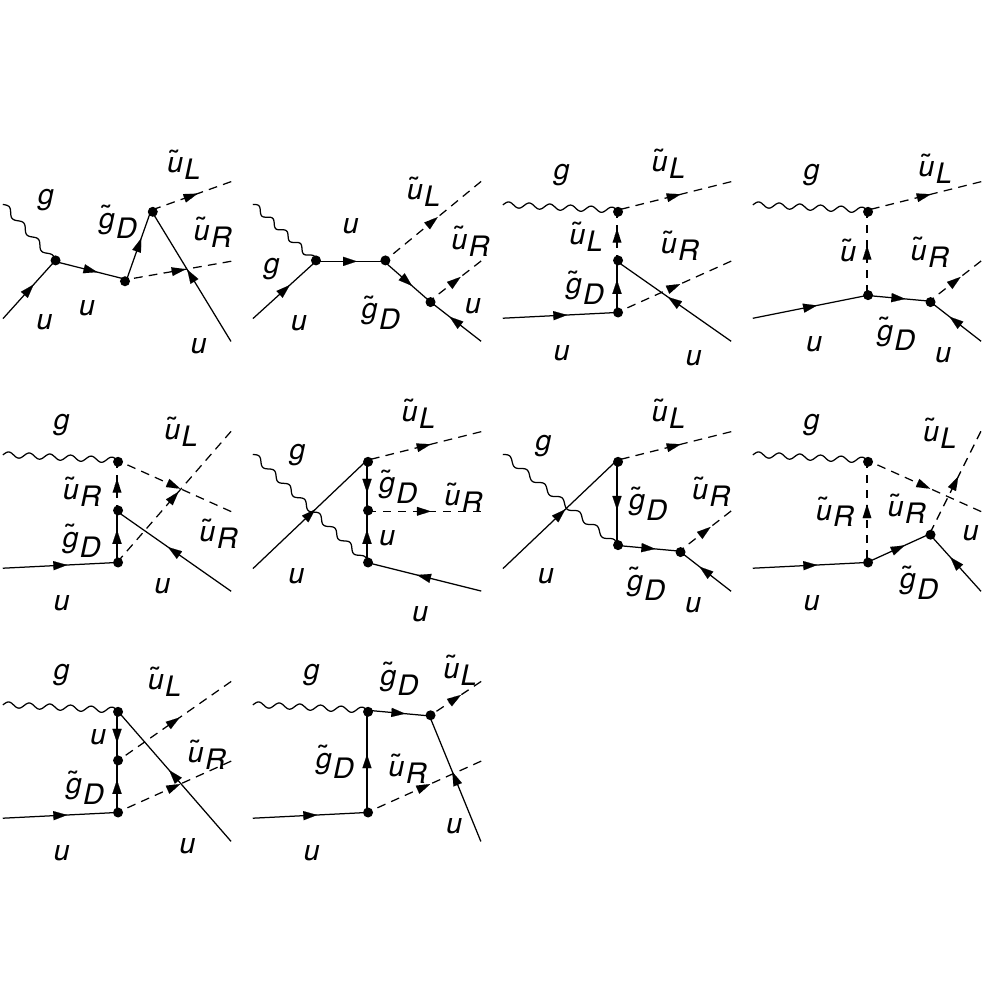}
  \caption{
    Feynman diagrams for $g u \to \tilde u_L \tilde u_R \bar u$ subprocess.
    For $m_{\tilde g_D} > m_{\tilde u_{L/R}}$, amplitude exhibit an on-shell singularity.
    \label{fig:gu_feynman_diagrams}
  }
\end{figure}

The calculation of the real emission corrections poses one more difficulty.
Figure \ref{fig:gu_feynman_diagrams} shows the diagrams for $g u \to \tilde u_L \tilde u_R \bar u$ process.
This process contains a contribution from the s-channel production of a gluino.
These kinds of effects, observed already in the SM for single-top production, require a careful definition of the process of interest.
The amplitude for diagrams in figure \ref{fig:gu_feynman_diagrams} can be split into resonant and non-resonant parts as
\begin{align}
  M_{tot} = M_{nr} + M_r .
\end{align}
For $m_{\tilde g_D} > m_{\tilde u_{L/R}}$, the $M_r$ becomes infinite when $(p_{\tilde g_D} + p_{\tilde u_{L/R}})^2 = m_{\tilde g_D}^2$.
We therefore employ the diagram removal (DR) procedure of Ref.~\cite{Frixione:2008yi}, where resonant diagrams are removed at the amplitude level.
As this procedure breaks gauge invariance, it requires a careful choice of the gauge for the external gluon. 
We choose a light-cone gauge, with the gauge vector $\eta$ set equal to the momentum of the second incoming parton.

\subsection{Results}
For numerical studies, we consider 2 MRSSM scenarios with a different gluino-squark mass hierarchy. 
Both points have common squark masses $m_{\tilde{q}} = 1.5$ TeV, $m_{O_A} = 5$ TeV, $m_t = 172$ GeV and differ only in the value of the gluino mass, with 1 TeV mass in the BMP1 and 2 TeV one in the BMP2.
The scalar sgluon mass $m_{O_S}$ is fixed by the eq.~\ref{eq:mass_relation}.
We perform all the calculations for the LHC running at center-of-mass energy of 13 TeV using \texttt{MMHT2014nlo68cl} PDFs interfaced through \texttt{LHAPDF6}.

The total, NLO cross section in the BMP1 is
\begin{align}
  \sigma_{\text{MRSSM}}^{\text{BMP1}}(p p \to \tilde u_L \tilde u_R) = 9.94312 \pm 0.00058 \text{ fb}.
  \label{bmp1}
\end{align}
This can be compared with the MSSM result obtained using \texttt{MadGraph5\_aMC@NLO}~\cite{Alwall:2014hca} and an NLO-capable UFO~\cite{Degrande:2011ua} model \texttt{SUSYQCD\_UFO} \cite{Degrande:2015vaa}, which is
\begin{align}
    \sigma_{\text{MSSM}}^{\text{BMP1}}(p p \to \tilde u_L \tilde u_R) & = 9.431 \pm 0.003 \text{ fb} .
\end{align}
The difference is of the order of 5\%, and results exclusively from the virtual part.

Similarly, for the BMP2 and after employing the diagram removal procedure, we get  
\begin{align}
  \sigma_{\text{MRSSM}}^{\text{BMP2}}(p p \to \tilde u_L \tilde u_R) =  3.01357 \pm 0.00020 \text{ fb}
  \label{bmp2}
\end{align}
versus the MSSM's
\begin{align}
\sigma_{\text{MSSM}}^{\text{BMP2}}(p p \to \tilde u_L \tilde u_R) & = 2.78114 \pm 0.00055 \text{ fb} ,
\end{align}
giving an 8\% difference.

Since the Born and the real matrix elements are identical for those processes in both models, it is not surprising that differences, though not negligible, do not exceed a few percent.

The results for the MRSSM are in full agreement with our second calculation based on \texttt{MadGraph5\_aMC@NLO} and \texttt{GoSam}~\cite{Cullen:2011ac,Cullen:2014yla}, interfaced using \texttt{BLHA2} interface \cite{Alioli:2013nda}.
However, we postpone the description of this method to further publication.

In the appendix, we give the decomposition of numbers in eqs~\ref{bmp1} and \ref{bmp2} into tree, virtual, soft and/or collinear and hard non-collinear contributions.
For phase space slicing parameters $\delta_s$ and $\delta_c$ we chose values $10^{-5}$ and $10^{-6}$, respectively.
We checked extensively that the final result is independent of precise numerical values of those (unphysical) parameters.
We also verified results for separate partonic channels between our \texttt{C++} and \texttt{MadGraph5\_aMC@NLO} implementations.

The above-mentioned \texttt{C++} code, containing $p p \to \tilde u_L \tilde u_R$ and $p p \to \tilde u_L \tilde u_L^*$ processes, will become public soon together with a more detailed description of both calculations.

%% file: conclusions.tex
\section{Conclusions}

The Minimal R-symmetric Supersymmetric Standard Model is a well-motivated and viable alternative to the MSSM.
Recent studies have shown that within the MRSSM framework one can accommodate the Higgs boson discovered at the LHC while staying in agreement with constraints from electroweak precision observables. 
Also, the first very promising studies of the dark matter sector were performed.

With respect to this progress, the study of the SQCD sector is lagging behind.
With that in mind, we have presented in this note our recent progress in that matter.
This included both the discussion of the recently derived limits on the sgluon mass from the LHC Run 2 data, as well as the calculation of the NLO SQCD corrections to the same-sign squark pair production.
To cross-check the latter results, we did two separate calculations, one based on analytic results encapsulated into a standalone \texttt{C++} code and the other using a well-established framework of \texttt{MadGraph5\_aMC@NLO}, which are in full agreement.

Our calculations showed a moderate difference of 5 - 10\% between MSSM and MRSSM for the considered benchmark points, which can be traced back to the difference in the virtual matrix element.

The developed \texttt{C++} code is already also capable of calculating the $p p \to \tilde u_L \tilde u_L^*$ process.
We plan to release the code to the public together with a description of the theoretical setup of the calculation.
This will be followed shorty by a phenomenological analysis.

%% file: acknowledgements.tex
\ack

I would like to thank Philip Diessner, Jan Kalinowski, Sebastian Liebschner and Dominik Stoeckinger, with whom the work on which this note is based was done.
Work supported in part by the German DFG grant STO 876/4-1, the Polish National Science Centre HARMONIA project under contract UMO-2015/18/M/ST2/00518 (2016-2019) and the PL-Grid Infrastructure.